\documentclass{article}

\usepackage{booktabs} 
\usepackage{graphicx}
\usepackage{tabularx}
\usepackage{amsmath}
\usepackage{url}
\usepackage{xcolor}
\usepackage{soul}

\title{SIFO: Secure Computational Infrastructure using FPGA Overlays}

\author{Xin Fang \\ Qualcomm, Inc.\\ Boxborough, MA \\
\tt{xinfang@qti.qualcomm.com}
\thanks{Xin Fang is currently affiliated with Qualcomm. The bulk of this research was done at Northeastern University.} \and
Stratis Ioannidis and Miriam Leeser\\
E.C.E.~Department, Northeastern University \\
\tt{ioannidis@ece.neu.edu} and 
\tt{mel@coe.neu.edu} \\
Boston, MA }

\begin{document}

\maketitle

\begin{abstract}

Secure Function Evaluation (SFE) has received recent attention due to the massive collection and mining of personal data, but remains impractical due to its large computational cost. Garbled Circuits (GC) is a protocol for implementing SFE which can evaluate any function that can be expressed as a Boolean circuit and obtain the result while keeping each party's input private. Recent advances have led to a surge of garbled circuit implementations in software for a variety of different tasks. However, these implementations are inefficient and therefore GC is not widely used, especially for large problems. This research investigates, implements and evaluates secure computation generation using a heterogeneous computing platform featuring FPGAs.  We have designed and implemented SIFO: Secure computational Infrastructure using FPGA Overlays.  Unlike traditional FPGA design, a coarse grained overlay architecture is adopted which supports mapping SFE problems that are too large to map to a single FPGA. Host tools provided include SFE problem generator, parser and automatic host code generation. Our design allows re-purposing an FPGA to evaluate different SFE tasks without the need for reprogramming, and fully explores the parallelism for any GC problem. Our system demonstrates an order of magnitude speedup compared with an existing software platform.
\end{abstract}

\section{Introduction}
\label{chap:intro}

The statistical analysis of data  collected from human subjects has a long history in empirical sciences such as medicine, sociology, and economics.  It has recently also become a ubiquitous practice among Internet companies, occurring presently at a massive and an unprecedented scale. Companies like Google, Netflix, and Amazon routinely monitor and mine a broad array of behavioral signals collected from their users, and monetize it through targeted advertising or personalized product recommendations. Behavioral data collection is therefore of considerable business value to online companies \cite{angwin2010web}; moreover, there are often benefits to society at large, as in aiding the detection of epidemics \cite{ramos2015google} or terrorist threats \cite{ressler2006social}, in assessing  news or product penetration~\cite{leskovec2009meme}, and in political online polling \cite{adamic2005political}. On the other hand, these practices have also given rise to privacy concerns and threats, documented extensively by researchers~\cite{kosinski2013private,salman:2013,blurme:2012,Narayanan:2008,Mislove:2010,Otterbacher:2010,Rao:2010} as well as the popular press \cite{angwin2010web,wortham2010facebook}. 

\subsection{Secure Function Evaluation}\label{sec:SFEintro}
This state of affairs gives rise to the following challenge:   given the benefits of mining behavioral data to both online companies and to society at large, 
is it possible to \emph{enable data mining practices without jeopardizing user privacy}? A series of recent research efforts  \cite{nikolaenko2013privacylinear,nikolaenko2013privacy,nayak2015graphsc,BEL11,du2004privacy,Huang2011} have attempted to address this issue through cryptographic means and, in particular, through secure function evaluation (SFE).  SFE allows an interested party to evaluate any desirable polynomial-time function over private data, while revealing only the answer and nothing else about the data. This offers a strong privacy guarantee: an entity executing a secure data-mining algorithm over user data learns only the final outcome of the computation, while the data  is never revealed to the entity. SFE can thus enable, e.g., a data analyst, a medical professional, or a statistician, to conduct a study of sensitive data without jeopardizing the privacy of the participants (online users, patients, etc.).

Any algorithm to be executed over amounts of data at the scale encountered in the above settings needs to be highly efficient and scalable. 
SFE over private data therefore poses a significant challenge, as it comes at a considerable additional computational cost compared to execution in the clear. Prior work has made positive steps in this direction, showing that  a variety of important data mining algorithms \cite{nikolaenko2013privacylinear,nikolaenko2013privacy,nayak2015graphsc} can be computed using Yao's Garbled Circuits (GCs)~\cite{yao1982protocols,Yao86} in a parallel fashion. The function to be evaluated is converted to a binary circuit which is ``garbled'' in such a way that an evaluator of the circuit learns only the values of its output gates. Execution of this circuit is subsequently parallelized, e.g., over threads \cite{nikolaenko2013privacy} or across a cluster of machines \cite{nayak2015graphsc}. Nevertheless, this approach to parallelization leaves much to be desired: for example, in \cite{nayak2015graphsc},
even under parallelization over 128 cores,  executing a typical data-mining algorithm like Matrix Factorization (MF) through SFE is of the order of $10^5$ slower compared to (parallel) execution in the clear. In practice, this means that applying MF to a dataset of 1M entries requires roughly 11 days under SFE, a time largely prohibitive for practical purposes. 

\subsection{FPGA Overlays}

There has been a surge of interest in FPGAs in the data center, as evidenced by a large number of systems that have recently become available.  Amazon is offering FPGA instances through Amazon Web services~\cite{amazonf1}, Microsoft has the Catapult system~\cite{caulfield2016cloud} and IBM offers cloud FPGA~\cite{ibm-fpga}.    
In this paper, we advocate leveraging hardware acceleration to tackle the  scalability and efficiency challenges inherent in SFE.  FPGAs are an excellent hardware platform  for the implementation of SFE primitives and, in particular, garbled circuits.
This is precisely because {FPGAs} are tailored to executing many low level operations in parallel. The types of operations encountered in garbled circuits (namely, garbling and un-garbling gates) fit this pattern precisely: they involve, e.g., a series of symmetric key encryptions, XOR{}s, and other well-defined primitive operations (see  Section~\ref{background}). Thus, an FPGA implementation of SFE benefits from both high speed evaluation and hardware-level parallelization. 

The amount of computation required to evaluate a garbled circuit for an  application at the usual data-mining scale cannot fit in a single FPGA. Thus, evaluating a function securely entails partitioning computations into  sub-tasks to be programmed and evaluated  over a single FPGA. A practical implementation therefore needs to allow repurposing an FPGA to quickly compute different SFEs or different sub-tasks of a larger SFE. For this reason, tailored approaches that are tied to the execution of a specific SFE structure, and require full reprogramming of an FPGA with each new execution, cannot be applied efficiently to the types of SFE problems we wish to address. 
To address these challenges, we propose a \emph{generic, reconfigurable implementation of SFE as a coarse-grained FPGA overlay architecture}.  As FPGAs have become more dense and capable of holding a large number of gate equivalents, there has been an increased interest in FPGA overlay architectures \cite{brant2012zuma,wiersema2014embedding,kapre2006packet,kapre2015hoplite,koch2013efficient,jain2015efficient,jain2016coarse}.  An FPGA overlay consists of two parts:  (1) a circuit design implemented on the FPGA fabric using the usual design flow, and (2) a user circuit mapped onto that overlay circuit.  Garbled circuits are excellent candidates for an FPGA overlay design. Precisely because components of a garbled circuit follow a generic structure, an overlay approach that does not reprogram FPGAs from scratch, but simply {\em reroutes} connections between elementary components (in our case, garbled AND and XOR gates) leads to important efficiency improvements.

\subsection{Contributions}
\label{contri}

This paper introduces SIFO: Secure computational Infrastructure using FPGA Overlays.  We make the following contributions:

\begin{itemize}

\item We provide a complete work flow to map {\em any} garbled circuit problem to garbled circuit overlay cells on an FPGA, including software (SFE problem generator, parser, and scheduler) and FPGA overlay circuit to accelerate the GC problem. 

\item  Our workflow and tools enable accelerating any garbled circuit operation without requiring knowledge of the underlying implementation. We integrate our implementation with FlexSC~\cite{flexsc} which uses ObliVM~\cite{liu2015oblivm} as the backend for any garbled circuit operation. In conjunction with our tools, each problem is analyzed and layers of operations that can be executed in parallel are extracted. The resulting circuit is then mapped to our FPGA overlay architecture for processing.

\item Our FPGA overlay architecture handles different parts of the same GC problem (if a problem is too large to fit in a single FPGA) as well as different GC problems without reprogramming.  The FPGA is programmed once for all garbled circuit problems. Wiring and instantiation are determined at execution time by the controller and the host. This overlay architecture is scalable and enables users to avoid the long design and compile time on FPGAs for new problems. The overhead for a new problem is very low, simply requiring the transfer of initial data and circuit information from host to device.  

\item We demonstrate the benefits of our approach by mapping a large number of circuit examples onto a heterogeneous computing platform featuring a Stratix V FPGA.  We tackle different aspects of performance bottlenecks and alleviate them.  This includes (a) investigating different numbers of FPGA overlay cells, (b) optimizing the host to FPGA communication via PCIe, and (c) managing on-chip block memory to minimize accesses to off-chip DDR memory.  We compare the performance of these improvements for various problems and show significant speed-up against the naive design and against a software implementation, ranging from 6.21 to 45.78 times faster than the latter.  Many of the optimizations presented can be applied to other FPGA projects as well.  

\end{itemize}

This journal paper represents an extension to our previously published research~\cite{fang2017secure}, which presented an implementation where the entire GC problem fit on a single FPGA, with all intermediate results fitting into on-chip memory (block RAM).  In the research presented here, we relax that constraint to significantly increase the size of problems supported. This introduces new challenges, since with GC, data is randomly accessed.  Our new implementation treats block RAM as user managed cache, and investigates how best to access data so that data fetching does not become a bottleneck.  

The remainder of this paper is structured as follows. Section 2 covers background information on garbled circuits as well as related work. The design methodology is presented in Section~\ref{arch}, which demonstrates the methodology of how we tackle the garbled circuit problem in a heterogeneous reconfigurable system, and how we alleviate bottlenecks in the system to improve overall performance. Experiments and corresponding results are presented in Section~\ref{chap:results}. Finally, we present our conclusions and future work. Material in this article are excerpted from the first author's PhD dissertation~\cite{fang2017privacy}.

\section{Background}
\label{background}

In this section, we introduce the relevant background on garbled circuits, including terminology and techniques. Related work on garbled circuit implementations is also discussed.

\subsection{Garbled Circuits}
\label{gc}

Our research accelerates Secure Function Evaluation (SFE), specifically Garbled Circuits (GC), using FPGAs. In this model there are two or more users with data which they wish to keep private, and a function to be evaluated over that data.  All parties know the function being evaluated and learn the outcome of the evaluation, but users do not reveal their data.  A canonical problem exemplifying SFE is the ``Millionaires' Problem'': two millionaires wish to know who is worth more without revealing their personal worth to each other.  

Garbled circuits were initially introduced by Yao~\cite{Yao86} for two users and has been extended to multiple users.  They rely on cryptographic primitives.  In the variant we study here (adapted from \cite{NPS99,nikolaenko2013privacy}), Yao's protocol runs between (a) a set of private input owners,  (b) an Evaluator,  who wishes to evaluate a function over the private inputs, and (c) a third party called the Garbler, that facilities and enables the secure computation.   

\begin{figure}[htb]
\begin{center}
\includegraphics[width=10 cm]{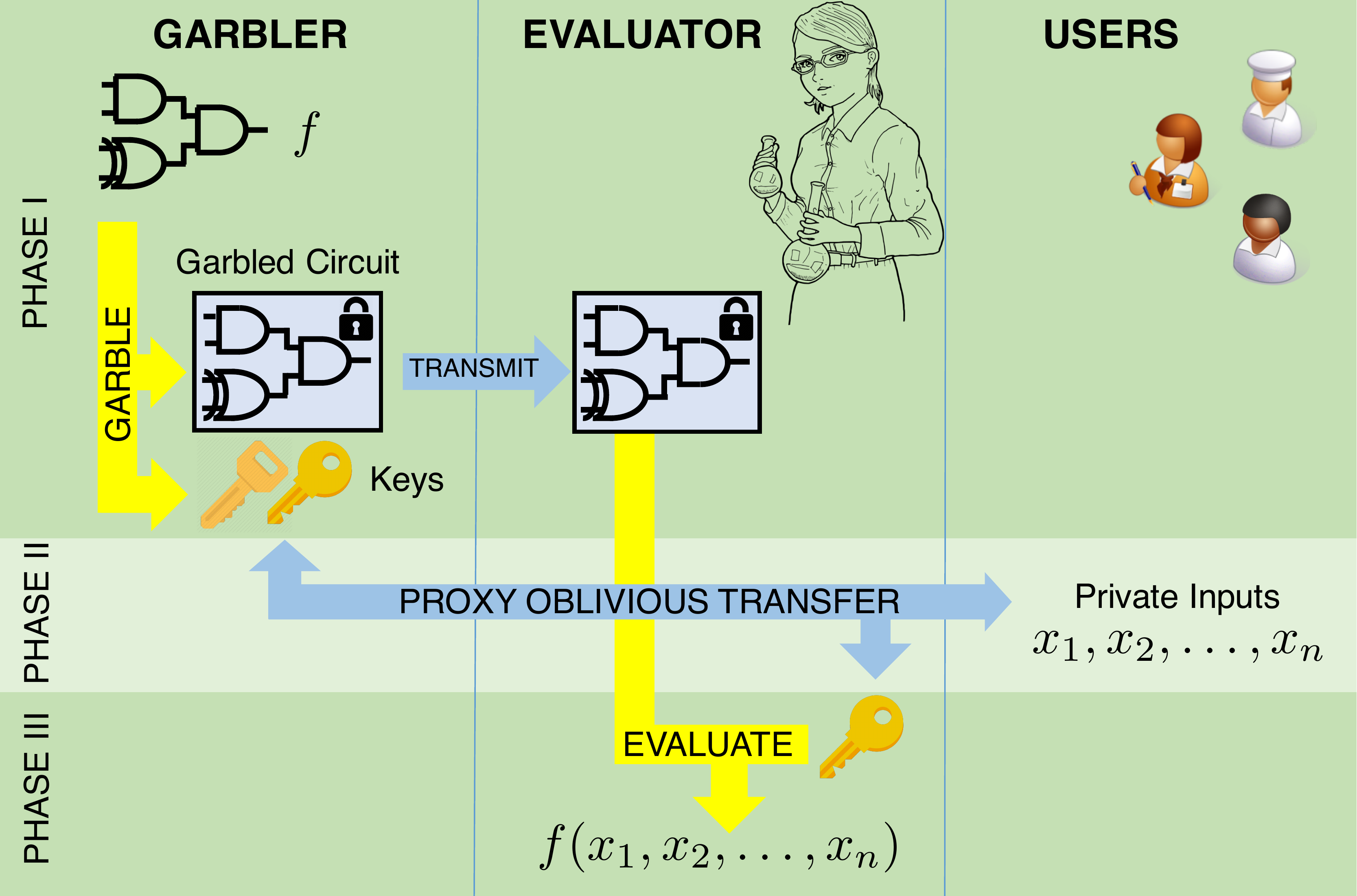}
\end{center}
\caption{Yao's Protocol Phases of Operation}
\label{fig:yaosprotocol}
\end{figure}

Garbled Circuits work for any problem that can be expressed as a Boolean circuit.  In our and many other implementations, this function is represented as a circuit made up of AND and XOR gates\footnote{Recall that AND and XOR gates form a complete basis for boolean circuits.}.  The Evaluator wishes to evaluate a function $f$, represented as a Boolean circuit of AND and XOR gates, over private user inputs $x_1,x_2,\ldots,x_n$. We break the problem into three phases, as shown in Fig.~\ref{fig:yaosprotocol}.  
In Phase I, the Garbler ``garbles'' each gate of the circuit, outputting (a) a ``garbled circuit,'' namely, the garbled representation of every gate in the circuit representing $f$, and (b) a set of keys, each corresponding to a possible value in the string representing the inputs $x_1,\ldots,x_n$. These values are shared with the Evaluator. In Phase II, through proxy oblivious transfer~\cite{naor2001efficient}, the Evaluator learns the keys corresponding to the true user inputs. In the final phase, the Evaluator uses the keys as input to the garbled circuit to evaluate the circuit, ungarbling the gates. At the conclusion of Phase III,  the Evaluator learns $f(x_1,\ldots,x_n).$  To ensure privacy of users' data and to protect against side channel attacks, both garbling and evaluation are run whenever user data changes.  Hence garbling is done as often as evaluation. 

\subsubsection{Garbling Phase}

A function to be evaluated is represented as a Boolean circuit 
consisting of AND and XOR gates.  In the garbling phase, each of these gates is garbled as described in this section.  Each gate is associated with three wires: two input wires and one output wire. At the beginning of the garbling phase, the Garbler associates two random strings, $k_{w_i}^0$ and $k_{w_i}^1$, with each wire $w_i$ in the circuit. Intuitively, each $k_{w_i}^b$ is an encoding of the bit-value $b\in\{0,1\}$ that the wire $w_i$ can take.  

\begin{figure}[htb]
\centering
\includegraphics[width=0.2\columnwidth]{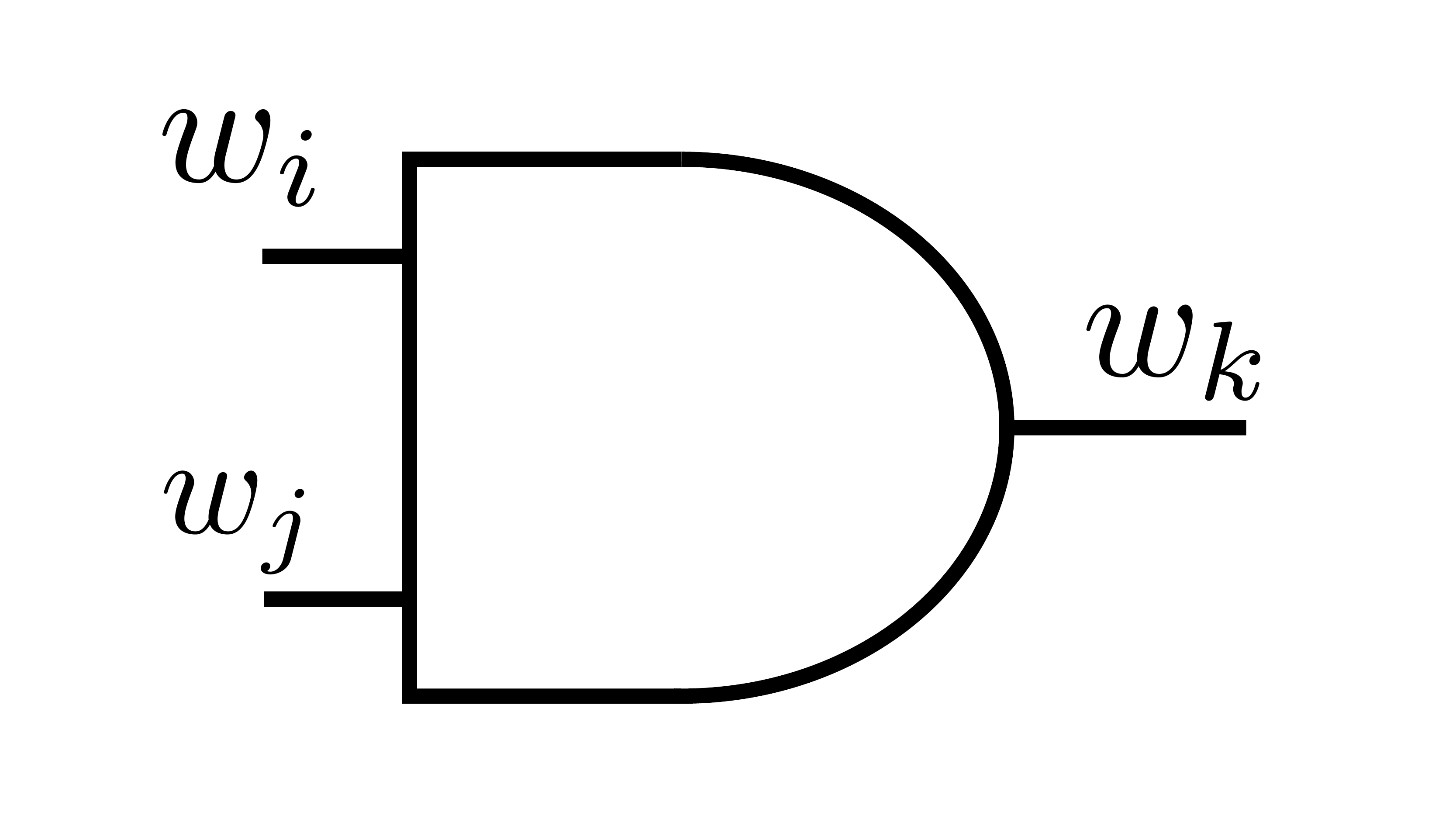}
\begin{small}
  \begin{tabular}{cccc}
    \hline\hline
    $b_i$ & $b_j$ & $f(b_i,b_j)$ & Garbled value\\
    \hline
    0 & 0 & 0 & $Enc_{(k_{w_i}^{0},k_{w_j}^{0},g)}(k_{w_k}^{0})$\\
    0 & 1 & 0 & $Enc_{(k_{w_i}^{0},k_{w_j}^{1},g)}(k_{w_k}^{0})$\\
    1 & 0 & 0 & $Enc_{(k_{w_i}^{1},k_{w_j}^{0},g)}(k_{w_k}^{0})$\\
    1 & 1 & 1 & $Enc_{(k_{w_i}^{1},k_{w_j}^{1},g)}(k_{w_k}^{1})$\\
    \hline\hline
  \end{tabular}
\end{small}
\caption{A Garbled AND Gate}\label{fig:andgate}
\end{figure}

We describe here how to garble an AND gate. The same principles can be applied to garble an XOR gate, using its respective truth table. We note however that, in practice,  XOR gates are handled via the Free XOR optimization \cite{freexor}, discussed in Section \ref{sec:optim}. A garbled AND gate is shown in Fig.~\ref{fig:andgate}.  For each AND gate $g$ where $g$ is the gate number, with input wires $(w_i,w_j)$ and output wire $w_k$, the Garbler computes the following four ciphertexts, one  for each pair of values $b_i,b_j \in \{0,1\}$:

\begin{align}
Enc_{(k_{w_i}^{b_i},k_{w_j}^{b_j},g)}(k_{w_k}^{g(b_i,b_j)})=SHA(k_{w_i}^{b_i}\|k_{w_j}^{b_j}\|g) \; \oplus \; k_{w_k}^{g(b_i,b_j)}
  \label{garbles}
\end{align}

Here SHA represents the hash function,  $\|$ indicates concatenation, $g$ is an identifier for the gate, and $\oplus$ is the XOR operation.  Note that each value $k$ on a wire is implemented with 80 bits in our implementation. 
The ``garbled'' gate is then represented by a random permutation of these four ciphertexts. 
Observe that, given the pair of keys $(k_{w_i}^{0},\allowbreak k_{w_j}^{1})$ it is possible to
successfully recover the key $k_{w_k}^{1}$ by decrypting
$c=Enc_{(k_{w_i}^{0},k_{w_j}^{1},g)}(k_{w_k}^{1})$ through\footnote{Note that the above encryption scheme is \emph{symmetric}, as Enc and Dec are the same function.}:
\begin{align}Dec_{(k_{w_i}^{0},k_{w_j}^{1},g)}(c) = SHA(k_{w_i}^{b_i}\|k_{w_j}^{b_j}\|g) \; \oplus \; c. \label{dec} \end{align}
On the other hand,  the other output wire 
key, namely $k_{w_k}^{0}$, cannot be recovered.  More generally, it is
worth noting that the knowledge of (a) the ciphertexts, and (b) keys $(k_{w_i}^{b_i},k_{w_j}^{b_j})$ for some inputs $b_i$ and $b_j$
yields \emph{only} the value of key $k_{w_k}^{g(b_i, b_j)}$;  no other input or
output keys of gate $g$ can be recovered.  Any Boolean function can be garbled in this manner, by first representing it in AND and XORs, and garbling each such gate.  

\subsubsection{Evaluation Phase}

The output of the garbling process is (a) the garbled gates, each comprising a random permutation of the four ciphertexts representing each gate, and  (b) the keys $(k_{w_i}^{0},k_{w_i}^{1})$ for every wire $w_i$ in the circuit. At the conclusion of the first phase, the Garbler sends this information for all garbled gates to the Evaluator. It also provides the correspondence between the garbled value and the real bit-value for the circuit-output wires (the outcome of the computation): if $w_k$ is a circuit-output wire, the pairs $(k_{w_k}^0,0)$ and $(k_{w_k}^1,1)$ are given to the Evaluator. 
To transfer the garbled values of the input wires, the Garbler engages in a proxy oblivious transfer with the Evaluator and the users, so that the Evaluator obliviously obtains the garbled-circuit input value keys $k_{w_i}^b$ corresponding to the actual bit $b$ of input wire $w_i$. 

Having the garbled inputs, the Evaluator can ``evaluate'' each gate, by decrypting each ciphertext of a gate in the first layer of the circuit by applying equation \eqref{dec}: only one of these decryptions will succeed\footnote{This can be detected, e.g., by appending a prefix of zeros to each key $k_{w_k}^b$, and checking if this prefix is present upon decryption.}, revealing the key corresponding to the output of this gate. Each output key revealed can subsequently be used to evaluate any gate that uses it as an input. Using the table mapping these keys to bits, the Evaluator can learn the final output.

\subsubsection{Optimization}\label{sec:optim}

Several improvements over the original Yao's protocol have been proposed, that lead to both computational and communication cost reductions. These include point-and-permute \cite{beaver1990round}, row reduction \cite{PSSW09}, and Free-XOR \cite{freexor} optimizations, all of which we implement in our design. Free-XOR in particular significantly reduces the computational cost of garbled XOR gates: XOR gates do not need to be encrypted and decrypted, as the XOR output wire key is computed through an XOR of the corresponding input keys. In addition, the free-XOR optimization fully eliminates communication between the Garbler and the Evaluator for XOR gates: no ciphertexts need to be communicated for these gates.  Our implementation takes advantage of all of these optimizations; as a result, the circuit for computing garbled AND gates differs slightly from the garbling algorithm outlined above.

\subsection{Related Work}
\label{rw}

Acceleration of garbled circuits is a hot research area in the SFE field. Researchers use different parallel models and hardware platforms to speed up execution. These platforms include FPGAs, CPUs, and GPUs.  

\subsubsection{FPGA and ASIC designs}

TinyGarble~\cite{songhori2015tinygarble} uses techniques from hardware design to implement GCs as sequential circuits and then optimizes these designs.  The circuits can be optimized to reduce the non-XOR operations using traditional high-level synthesis tools and simulation. 
 
The offline circuit synthesis will provide a ready-to-use circuit description for any garbled circuit problem.  The resulting designs are customized for each problem; thus for each new problem a new circuit must be generated.  In addition, their results describe simulations, but no actual hardware implementation. 
\cite{jarvinen2010garbled,jarvinen2010embedded} describe the first FPGA implementations of GC. In both these implementations, there is limited parallelism to allow garbling to happen in a small footprint. In~\cite{jarvinen2010garbled}, two FPGA-based prototypes are described, a system-on-chip with access to a single hardware cryptographic accelerator core, and a stand-alone hardware implementation targeting ASICs.  In~\cite{jarvinen2010embedded} the authors use a non-standard garbling technique in order to reduce communication.  Our approach uses standard GC techniques as implemented in popular software implementations. 
In addition, our architecture aims to reduce the computational cost of garbling by using much more parallelism than these early FPGA implementations. For starters, we implement four SHA cores in hardware for each garbled AND gate.
In addition, we implement as many garbled AND gates as we can keep busy at the same time, and implement garbled circuits directly on top of an efficient overlay, which eliminates the need to recompile the hardware for every new user problem.  With MAXelerator~\cite{hussain2018maxelerator} the authors implement a very efficient garbling of matrix multiplication in FPGAs.  While their design is more efficient for matrix multiplication, ours is more general purpose and supports any problem that a user may wish to garble.  

\subsubsection{CPU}

One approach to accelerating GC on CPUs is to provide instructions that support encryption to speed up the base operations.  
JustGarble~\cite{bellare2013efficient} shows that using AES-NI (Advanced Encryption Standard New Instruction), circuits can be garbled and evaluated  faster than using traditional instructions. Intel AES-NI is a new encryption instruction set that improves AES operations in the Intel Xeon processor family.  
Others have proposed a 32-bit MIPS architecture specifically implemented with instructions to support SFE.  GarbledCPU~\cite{songhori2016garbledcpu} is a MIPS-based general-purpose sequential processor which enables the high-level description of garbled circuits in hardware. Problems to be evaluated securely are compiled to MIPS assembler and then run securely on their garbled MIPS processor. The goal of this project is to fabricate the MIPS core; FPGAs are used for prototyping the design. Using MIPS assembly code to represent the problem being evaluated alleviates the problem of lengthy FPGA place and route cycles.  However, the availability of this specialized hardware is likely to be limited.  Our approach introduces more parallelism than either of these CPU approaches, as we implement many hashing cores in parallel.  In addition, through an overlay, we can rapidly switch between problems.  

\subsubsection{GPUs}

Researchers have used GPUs for hardware implementations of garbled circuits. Fastplay~\cite{pu2011fastplay} uses a GPU architecture to accelerate garbling arithmetic operations and achieves a 35 to 40x improvement over a serial implementation.   
Fredericksen et al. \cite{frederiksen2014faster} implement a protocol based on cut-and-choose of garbled circuits for malicious situation using GPUs. Husted et al. \cite{husted2013gpu} implement free-XOR, pipeline, and OT extension on GPUs which exploit some of the parallel nature of these tasks. They report on the difference between implementations on Single Instruction Multiple Data (SIMD) architecture of GPUs and on Multiple Instruction Multiple Data (MIMD) architectures for multi-core CPUs.  They also comment on the difficulty of comparing different implementations.  Husted assumes a malicious adversary, and thus implement $k$ different versions of a Garbled Circuit which gives them increased parallelism.  We assume an ``honest but curious'' adversary, which results in less parallelism.  

\subsubsection{Summary} Our approach, SIFO, differs from prior art with respect to  (a) the level of parallelism implemented, (b) the ability to support any user problem, and (c) the ease to change between problems without requiring regeneration of the FPGA circuit.

\section{System Design Methodology}
\label{arch}

Our approach implements a coarse-grained overlay architecture to accelerate GC problems.  Garbled AND and XOR gates are implemented on an FPGA along with memory and control for support.  Software tools support the mapping of different garbled circuit problems onto this overlay architecture and leverage the interaction between hardware and software while maintaining small communication and memory access overhead.  We describe the hardware architecture (Sec.~\ref{hwarch}), software structure (Sec.~\ref{swstru}), and discuss why an overlay architecture is needed.   We conclude the section with a discussion of  optimizations implemented for performance improvement.

To demonstrate the utility of FPGAs in the datacenter for accelerating GC, we start with circuits generated from  FlexSC based on ObliVM~\cite{liu2015oblivm}.  FlexSC is a software framework that allows developers without any cryptography expertise to convert algorithms expressed in a high-level language to GC. FlexSC generates a gate netlist of the problem to be garbled, where gates are restricted to AND and XOR gates.  This research takes this  netlist and processes it on an FPGA and compares it the same processing done by FlexSC on a CPU.  In this paper we focus on garbling. Our recent results~\cite{hpec2019gc} show that garbling takes up about two thirds of the total run time and is thus the bottleneck in our overall design.  In this paper we focus on garbling;  accelerating evaluation will be addressed in future work.

The overall process starts from user data and a problem to be garbled.  The steps required are generating the netlist for the garbled circuit, mapping that netlist onto implementations of AND and XOR gates, generating the garble tables for the evaluator, and then transmitting the table for each AND gate to the evaluator.  The evaluator receives data inputs from the users via oblivious transfer.  In the process of garbling, we use FLEXSC to generate the netlist and use the wire numbers from that netlist as the memory locations for each wire.   Software that runs on the host processor does layer abstraction to process a circuit in breadth first order, assigns AND operations in the garbled circuit to specific AND gates on the FPGA. Hence, what is communicated to the FPGA is wire IDs and gate numbers for garbling.  The garble tables are transferred back to the host processor for transfer to the evaluator.  This flow is shown in Fig.~\ref{fig:systemgc} and described more completely in this section.

\subsection{Hardware Architecture}
\label{hwarch}

\subsubsection{gAND and gXOR Overlay Cells}

\begin{figure}[htb]
\begin{center}
\includegraphics[width=10 cm]{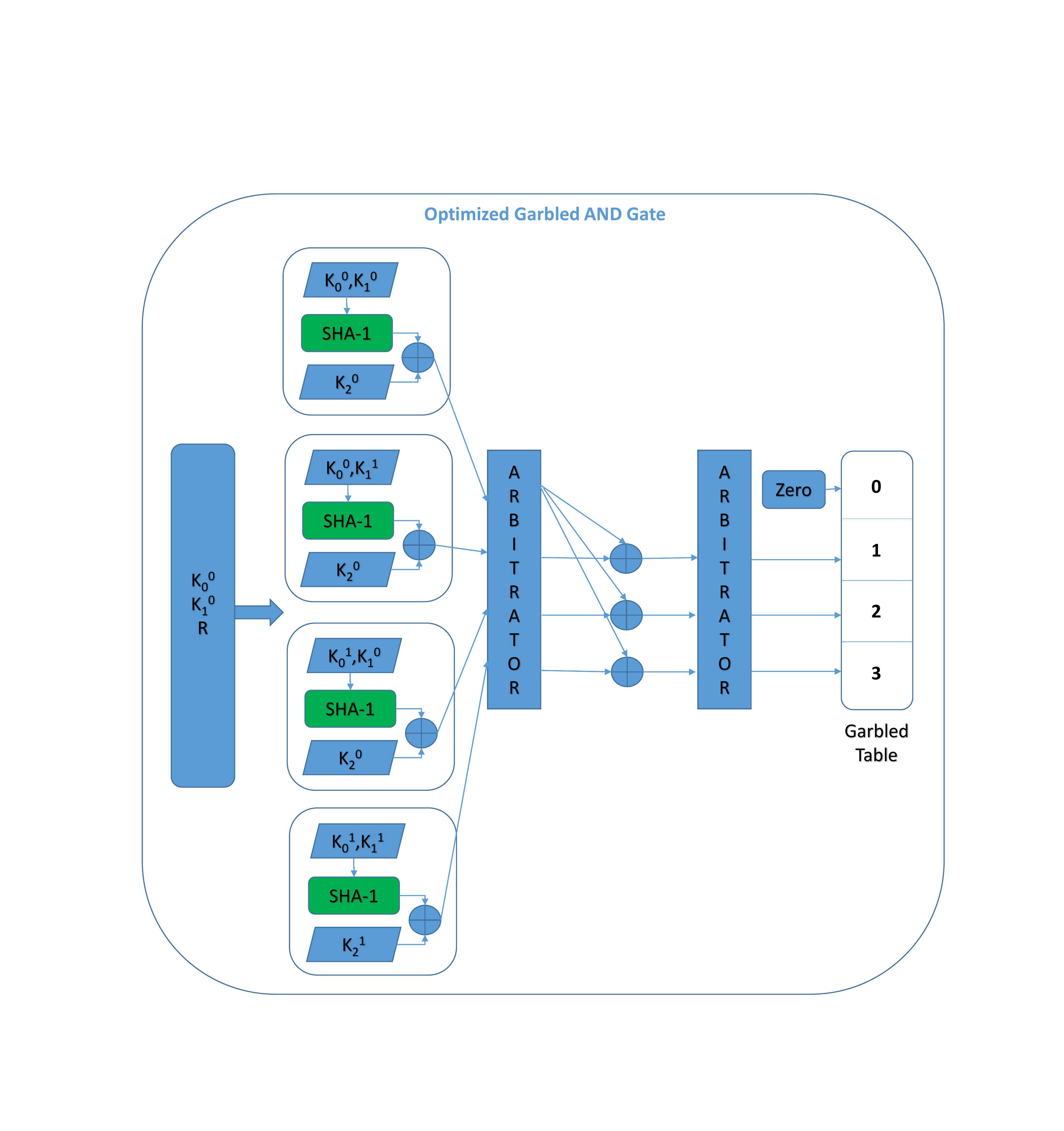}
\end{center}
\caption{Optimized Garbled AND Overlay Cell}
\label{fig:garbledopt}
\end{figure}

The garbled AND cells required for garbled circuit generation are much more complicated than single bit operations. To emphasize this fact, we refer to them as {\em gAND} in the reminder of this paper.  Each wire of the gAND is represented with 80 bits. A basic garbling AND operation implements the functionality described in Section~\ref{gc}.  The design we use, shown in Fig.~\ref{fig:garbledopt}, implements the row-reduction~\cite{PSSW09} and ``free''-XOR \cite{freexor} optimizations. Each line of the truth table is implemented according to Eq.~\ref{garbles}. This implementation requires four Secure Hash Algorithm (SHA) 1 cores, although only three output values need to be transmitted to the evaluator. $K_0^0$, $K_1^0$ are two garbled values representing the value $0$ on the wire for a gAND operation. R is a global variable based on which the cipher can get the garbled value represented by all zeros. For any wire i, $K_i^0 \oplus K_i^1 = R$.  
The implementation still uses four SHA-1 primitives which run in parallel; however only three values in the garbling table need to be stored; reducing the size of the garble table by 25\%.  Since all values in the garble table need to be transmitted to the host, and later to the evaluator, this optimization also results in a 25\% saving in the amount of data that needs to be transmitted for each gAND gate.   The implementation includes two arbitrators for the four outputs of the SHA-1 operations. The first arbitrator decides the sequence of the result and picks one of them to XOR with the other three. The second arbitrator rearranges the sequence of those three values and stores them in the garble table. Note that these arbitrators do not introduce any latency to the system. The latency of a gAND gate is 82 cycles, which is determined by the latency of the SHA-1 core.  The implementation, based on an open source core~\cite{fpga-shacore}, uses 512-bit values derived from the garbled inputs and additional information. gAND requires 82 clock cycles on the FPGA and uses 3070 ALMs and 3750 one bit registers on our target hardware, a Stratix V FPGA.  

There are several things to note about this implementation. First, SHA-1 is known to be vulnerable; however, since new keys are generated for every new problem and new set of inputs, this is not a concern in the context of GC.  A user who wants a stronger privacy guarantee can replace the SHA-1 cores with AES or another cryptographic primitive.  This may reduce the performance in our implementation as the number of cores that can be implemented in parallel could be reduced.  Second, other optimizations have been introduced, most importantly the half-AND gate~\cite{zahur2015two}, which reduces the amount of data that needs to be transmitted between garbler and evaluator.  This optimization will not accelerate garbling, the focus of this paper, but will reduce communication costs. These and other optimizations can easily be introduced into the design of gAND and will be considered in the future.  Note that SHA was chosen in order for us to compare our performance directly to a widely used software implementation, FlexSC~\cite{flexsc}.

The Garbled Circuit XOR overlay cell (referred to as gXOR) benefits from the  ``free'' XOR protocol \cite{freexor}. A free gXOR gate consists of 80-bit plaintext XOR operations. For any garbled circuit operation, it is guaranteed that using the free XOR approach will have the same privacy guarantees as using standard cryptographic primitives. This optimization means that gXOR is both much smaller and much faster than gAND.  Note that the gXOR gate is combinational and thus has no latency.  Note that the time to garble an XOR operation is less than the time to transfer the input and output wire information from the host processor.  However, it is still advantageous to do this in FPGA hardware since the input and output garbled values stay local to the FPGA and would otherwise have to be communicated back to the host. 

\subsubsection{FPGA Overlay Architecture}
\label{overlay}

\begin{figure}[htb]
\begin{center}
\includegraphics[width=10 cm]{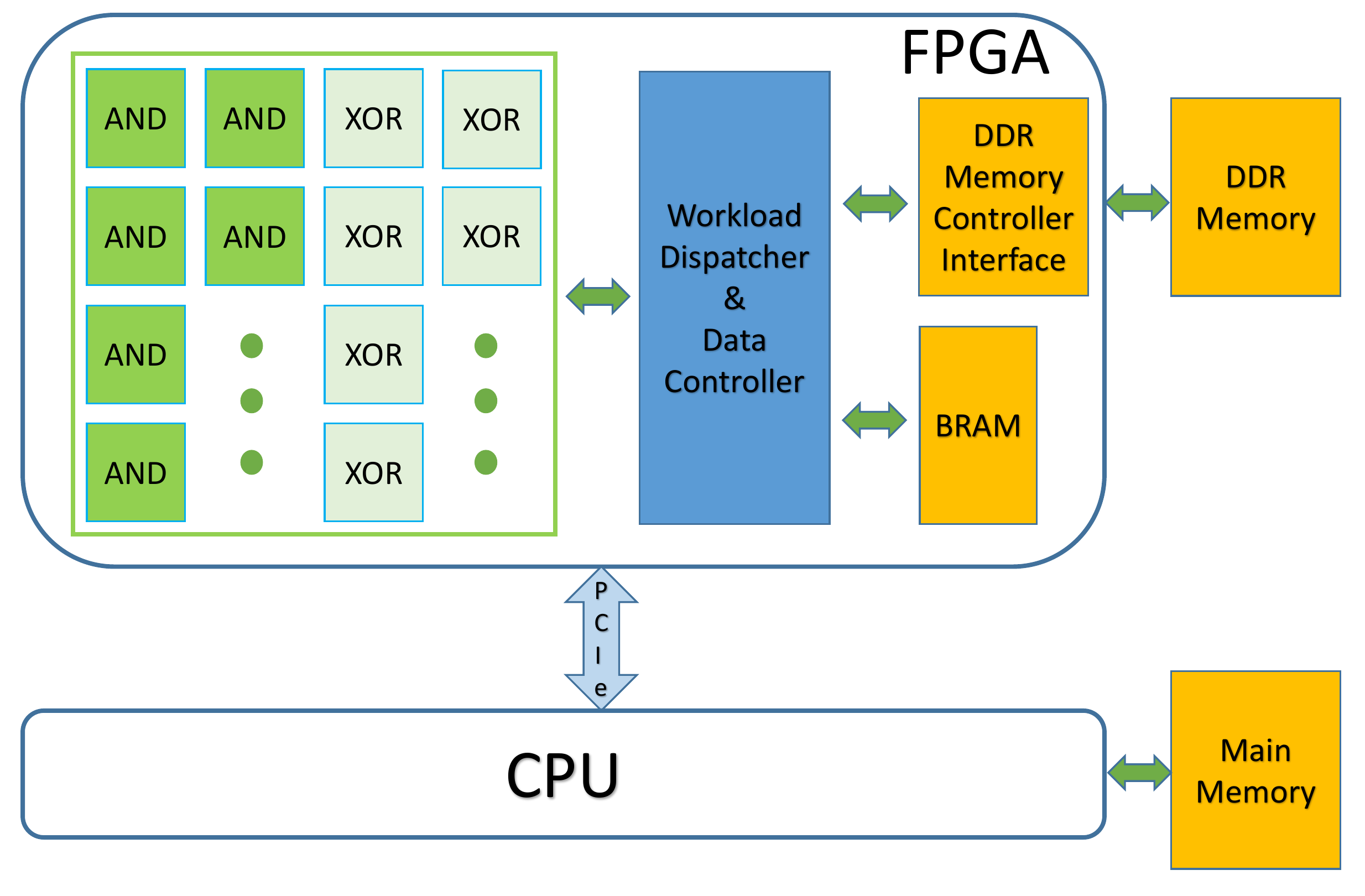}
\end{center}
\caption{Hardware Architecture}
\label{fig:hybridmem}
\end{figure}

Fig.~\ref{fig:hybridmem} shows the overlay architecture we use for garbled circuit acceleration. This architecture includes the gAND and gXOR circuits described above, a Workload Dispatcher and Data Controller (described below), Block RAM that is used as an on-chip cache and a DDR memory interface for accessing the main memory for the problem being garbled.
An architectural decision in our overlay design is how many gAND and gXOR gates to instantiate.  We experimented with different numbers of AND and XOR overlay cell combinations; the results are presented in Sec.~\ref{chap:results}.

\subsubsection{Workload Dispatcher and Data Controller}

\begin{figure}
\begin{center}
\includegraphics[width=10 cm]{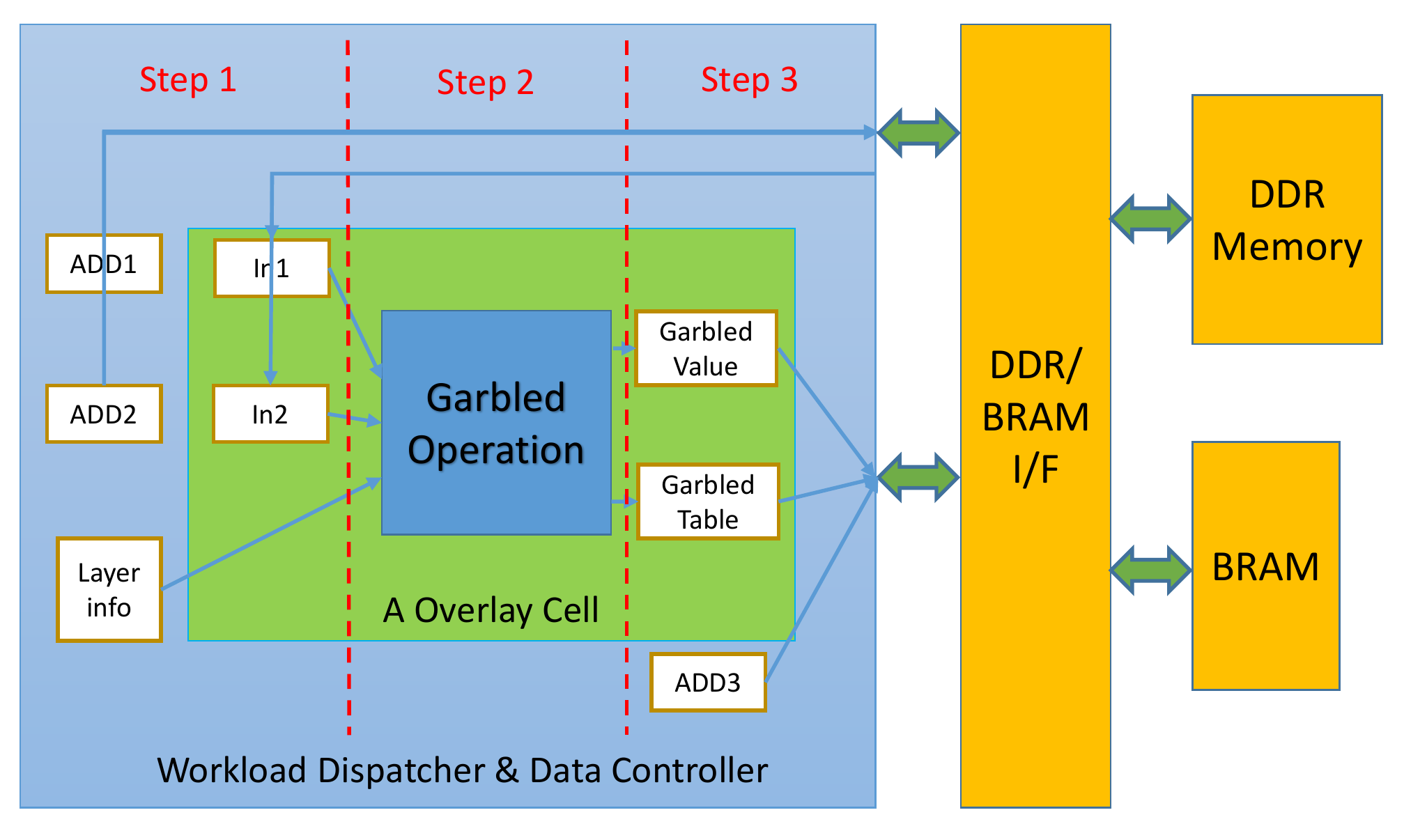}
\end{center}
\caption{Workload Dispatcher and Data Controller Timing information}
\label{fig:cont_time}
\end{figure}

The workload dispatcher and data controller is responsible for fetching garbled values from memory based on input and output addresses, delivering input data to the correct gAND or gXOR overlay cell, and writing back the results to memory after each operation. The timing of the entire system is determined by the states in this module. 
Fig.~\ref{fig:cont_time} shows the timing information of the workload dispatcher and data controller, which implements the following steps: (1) determining the type of the next batch of operations sent from the host, (2) reading input values from memory and forwarding them to the correct overlay cell, and (3) writing the output result back to the corresponding location in memory.  Our design uses both on-chip block RAM (BRAM) and off-chip DDR memory.  
States in the FSM are added depending on the type of memory accessed.  An entry using block RAM for storage will wait for only one clock cycle for a read or write; however, for DDR memory, the memory access operation has variable latency and is not finished until the ``complete'' flag is raised.  One of the challenges of GC is that memory locations are accessed in random order, hence timing and organization of memory is complicated.  We currently use on-chip BRAM as a cache for values that would otherwise be stored in off-chip memory.  A detailed discussion of different memory optimizations is presented in Sec.~\ref{chap:results}.  The on-chip BRAM is organized with a single read port and single write port with 108-bit data, of which 80 bits are used. 6.75 Mbits of BRAM is used in this design to store values.  Multiple BRAMs on the chip can be accessed in parallel.  However, the random access nature of memory accesses makes it challenging to take full advantage of this feature.  The on-board memory is accessed using 512 bit reads and writes, and four garbled values are accessed in one data word.  Two data ports are available in parallel.  The port widths are dictated by the architecture of the Gidel ProceV board.  

\subsection {Why use an Overlay Architecture?}

\begin{table}[htb]
\centering
\begin{scriptsize}
\caption{Gate Information for Problems}{
\begin{tabular}{lcccccccc}
\hline
Problem & Layers & Input Wires & Output Wires  & ANDs & XORs & Gates  & \# Reprogram \\
\hline
6-bit adder  & 17 &12 &6 & 6 & 24  & 30  &  1 \\
\hline
10-bit HD    & 22 &20 &10 & 20 & 90 & 110 & 1 \\
\hline
30-bit HD    & 27 &60 &30 & 60 & 270 & 330  & 6 \\
\hline
50-bit HD  & 32  &100 &50 & 100 & 450  & 550 & 10 \\
\hline
8-bit mult & 57 &16 &16 & 120  & 352  & 472 & 12 \\
\hline
16-bit mult & 121 &32 &32 & 496 & 1472 & 1968 & 50 \\
\hline
32-bit mult & 249 & 64 & 64 & 2016 & 6016 & 8032 & 201 \\
\hline
64-bit mult & 505 &128 & 128 & 8128 & 24320 & 32448 & 813 \\
\hline
10 4-bit sorting & 278  &40 & 40 & 848  & 4638 & 5486  & 85 \\
\hline
$5 \times 5$ 4-bit m\_mult  & 25 & 100 & 200 & 3900 & 11600 & 15500 & 390 \\
\hline
$10 \times 10$ 4-bit m\_mult & 27  &400 &800  & 7526  & 22489  & 30015 & 753 \\
\hline
$5\times 5$ 8-bit m\_mult   & 57 & 200 & 400  & 15800  & 47200  & 63000  & 1580 \\
\hline
$10 \times 10$ 8-bit m\_mult & 57  &800 &1600  & 127200    & 380800    & 508000 & 12720 \\
\hline
$20 \times 20$ 4-bit m\_mult & 37  &1600 &3200  & 254400    & 761600    & 1016000 & 25440  \\
\hline
\end{tabular}}
\label{gateinfo-motiv}
\end{scriptsize}
\end{table}

The examples used in Sec.~\ref{chap:results} are shown in Table~\ref{gateinfo-motiv}.  These problems  are addition, Hamming Distance (HD), multiplication, sorting,  and matrix multiplication. We also analyze scalability of our design by testing several different sizes of these problems. Note that FlexSC tries to maximize the number of XOR gates uses as XOR gates are much less computationally expensive to implement with the free XOR optimization~\cite{freexor}.  Thus the percentage of gates that are AND gates never exceeds 26\% in our examples.   

\begin{table}[htb]
\centering
\caption{Problem Switching Time}{
\begin{tabular}{cc|c}
\hline
\multicolumn{2}{c|}{Our Workflow} & Traditional Workflow \\
\hline
Hardware Architecture & Software Generation & Hardware Design\\
\hline
One Time Compile  & Minutes\  & Every problem \\
less than one hour & & minutes to hours\\
\hline
\end{tabular}}
\label{tab:switchtime}
\end{table}

The last column of Table~\ref{gateinfo-motiv} shows the number of times the FPGA would need to be reprogrammed assuming layers are processed one at a time and 10 garbled AND gates are implemented on the FPGA.  We assume as many XORs as needed can be accommodated, as XORs take much less time and space to process, as discussed above.   These examples motivate the need for an overlay architecture.  

The FPGA architecture is implemented as a coarse grained overlay circuit with a sea of gates approach, where the implemented gates are gAND and gXOR.  Using this architecture, any user problem can be mapped to the garbling hardware with no need to reprogram the FPGA, and the results are available rapidly.  This is in contrast to  the traditional FPGA design workflow which would require synthesis, place and route as well as downloading a new bitstream for each new problem.  In our approach, all that is needed is to program the FPGA ahead of time and generate the software host code for each problem. The overlay architecture can be reused without recompilation, while the traditional FPGA workflow has to go through the entire tool flow.
The traditional approach is infeasible for garbling large problems as many recompilations would be required for a single problem.  Each compilation can take minutes to hours. Using the overlay architecture, we compile the hardware once, and generate the software in minutes.  Let's assume that we used a very efficient program, such as TinyGarble~\cite{songhori2015tinygarble}, to generate each instance of a problem.  Lets further assume that TinyGarble can fit a design with 100 garbled AND cores on an FPGA, i.e.\ it is ten times more efficient in hardware usage compared to our approach.  To handle multiple problems, such as those in a data center setting, each new problem would need to be generated, placed and routed, and this takes on the order of tens of minutes.  Hence, our approach is more efficient even for those problems that fit entirely on one FPGA.  For large problems, such as the larger matrix multiply problems in Fig.~\ref{gateinfo-motiv}, the FPGA would have to be reprogrammed more than a hundred times, a process that would require hours. The overlay approach provides an architecture that maps different designs to the FPGA without requiring reprogramming.  Thus the end-to-end run time of an application with FPGAs is faster than the end-to-end run time using FlexSC, as presented in Sec.~\ref{chap:results}.  We summarize this discussion in Table~\ref{tab:switchtime}.

\subsection{Software Workflow}
\label{swstru}

This section discusses the software workflow including problem generation, problem parsing, layer extraction, and code generation.

\subsubsection{Problem Generation and Validation}

Our design makes use of our GC Overlay architecture, SIFO, in a way that is seamless for a user of FlexSC.  
FlexSC, based on ObliVM~\cite{liu2015oblivm}, is a software framework that allows developers without any cryptography expertise to convert algorithms expressed in a high-level language to GC. 
We modify FlexSC to output the netlist for a garbled circuit problem. This research extends FlexSC by taking the netlist, consisting of gAND and gXOR gates, and processing it on an FPGA. We use the same optimizations as FlexSC, namely free XOR~\cite{Kolesnikov2008} and row reduction~\cite{PSSW09}. 
In its normal operation, FlexSC outputs the results of garbling each gate; we use these values for verification.  Note that input values are random and generated for each new computation.  These are generated on the host and used for both the FlexSC and FPGA versions to ensure consistent results. The speed and validity of results can thus be easily compared.  

The netlist generated by FlexSC is {\em garbled} in breadth first order.  To support this, we generate layer information, and separate each layer into a ``batch'' of operations, where each batch represents the number of gates that can be garbled in parallel on the FPGA, and is implementation specific.  As the netlist can be quite large, it may require many batches to garble a single layer. A typical Boolean gate generated from FlexSC has the form: wire ID1 AND/XOR wire ID2 = wire ID3. We use WireIDs as memory addresses; intermediate data from garbling needs to be stored.  We use both on-chip and on-board memory for this purpose.

\begin{figure}
\begin{center}
\includegraphics[width=10 cm]{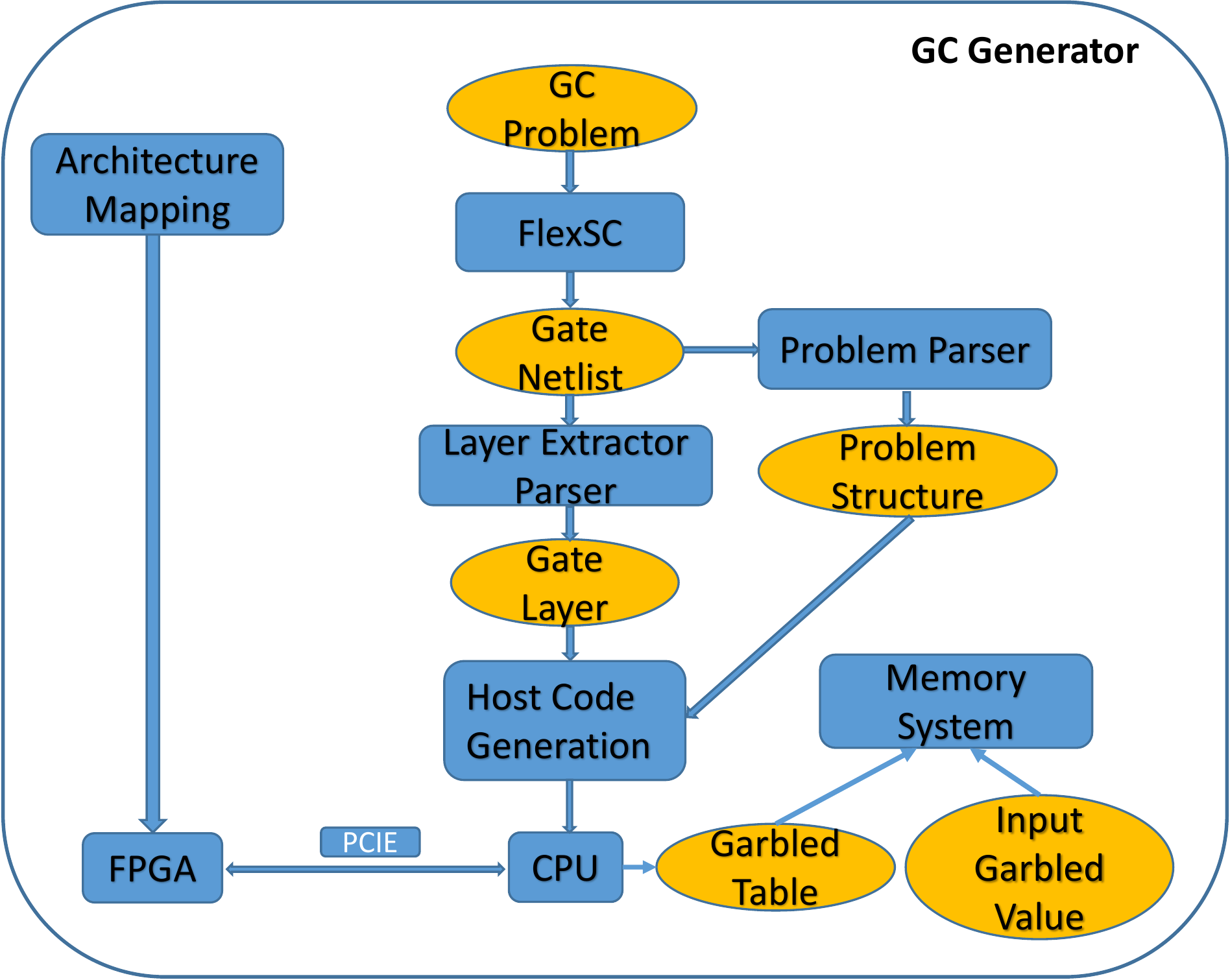}
\end{center}
\caption{Garbled Circuit Generator:  Hardware and Software}
\label{fig:systemgc}
\end{figure}

Our hardware consists of an FPGA board connected to a PC via PCIe.  
Fig.~\ref{fig:systemgc} highlights the workflow for garbling a circuit that involves both software running on a CPU (on the right of the figure) and FPGA design (on the left).  

The host transfers the information to the FPGA for processing.  We automatically generate the host code for each problem through our tools.
The host code for a user problem is responsible for initial data transmission, assigning gates in the problem to specific gAND or gXOR instances on the FPGA, and allocating the output of each garbling operation to memory.  

\begin{figure}
\begin{center}
\includegraphics[width=12 cm]{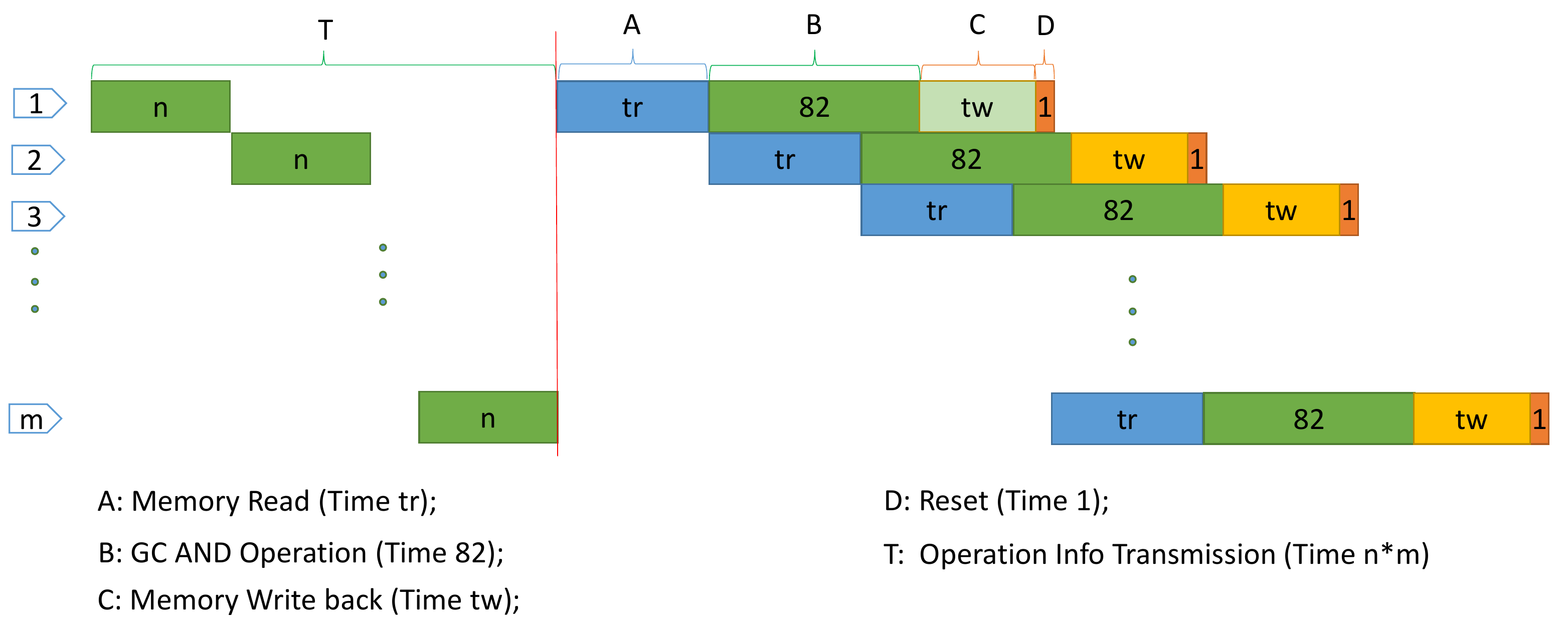}
\end{center}
\caption{CPU and FPGA Communication without Overlap}
\label{fig:coarse}
\end{figure}

Fig.~\ref{fig:coarse} shows the timeline for one batch of Boolean operations for the workload dispatcher and data controller, assuming that all information is transferred from the host before the batch begins operation.  This is improved on by overlapping communication and computation as described in the optimization section.

\subsubsection{Problem Parser and Layer Extractor}

The problem parser analyzes the generated gate netlist From FlexSC for gate, wire and layer information. 
The output consists of the total number of wires representing the total number of memory locations required, and, for each AND gate and XOR gate, the addresses that correspond to the input and output wires. Other output includes information for separating wires into different groups (on-chip or on-board) which is used in implementing the hardware memory hierarchy.  We consider several different approaches for using on-chip memory effectively. Analysis of results for sample problems is presented in Section~\ref{analysis}.  

We process gates in breadth first order.  The netlist generated from FlexSC is fed to a layer extractor which extracts each layer of the circuit that can be garbled in parallel. We also identify the primary input values whose wire ID is not the output of any gate.
Layer extraction identifies the AND and XOR gates that can be processed at the same time. For most problems, an entire layer will not fit onto the implemented FPGA overlay architecture.  Thus, a single layer may take several rounds.  We refer to the number of gates that map directly onto the FPGA as a batch.  Each operation in a batch is assigned a gate ID that corresponds to the gate it uses in HW.  
Wire numbers for input and output wires are used as the addresses in memory where input and output values are stored.  The processor assigns wire IDs and gate IDs and transmits this information to the FPGA.  At the end of a batch, the processor transmits the next batch of information, and continues until the circuit is fully garbled. Within a batch, all gates belong to the same layer of the circuit. Note that the amount of information transferred from host to FPGA minimal.  The data remains on the FPGA; only memory addresses and gate IDs are transmitted.  

\subsubsection{Host Code Generation}

We developed the tools to automatically generate the host code based on any garbled circuit operation. The input is the layer information from the layer extractor. 
This tool generates the batches and assigns wire IDs and gate IDs for different problems. Initial input data is generated and sent to the FPGA.  For very large problems the host code separates the main function into groups of smaller problems to avoid exceeding the heap size allocated for a problem.  The tools support debug mode, as well as different allocation policies for memory, which are discussed in more detail below.  More details can be found in \cite{fang2017privacy}.

\subsection{Optimizations}
\label{optim}

There are two major sources of bottlenecks in our design.  The first is transferring data over PCIe.  The second is the delay in accessing on-board memory.  In this section we address optimizations to the design that mitigate both of these bottlenecks.  

\subsubsection{PCIe Communication and FPGA Memory}
\label{pcieorg}

The first few optimizations target improving communications over the PCIe bus.  In our implementation, for each gate, the location of the input and output wire values and the gate type:  AND or XOR needs to be communicated.  Since the circuits representing problems to be garbled are large, this information is transferred as a batch of operations at a time.  

\begin{figure}[htb]
\begin{center}
\includegraphics[width=12 cm]{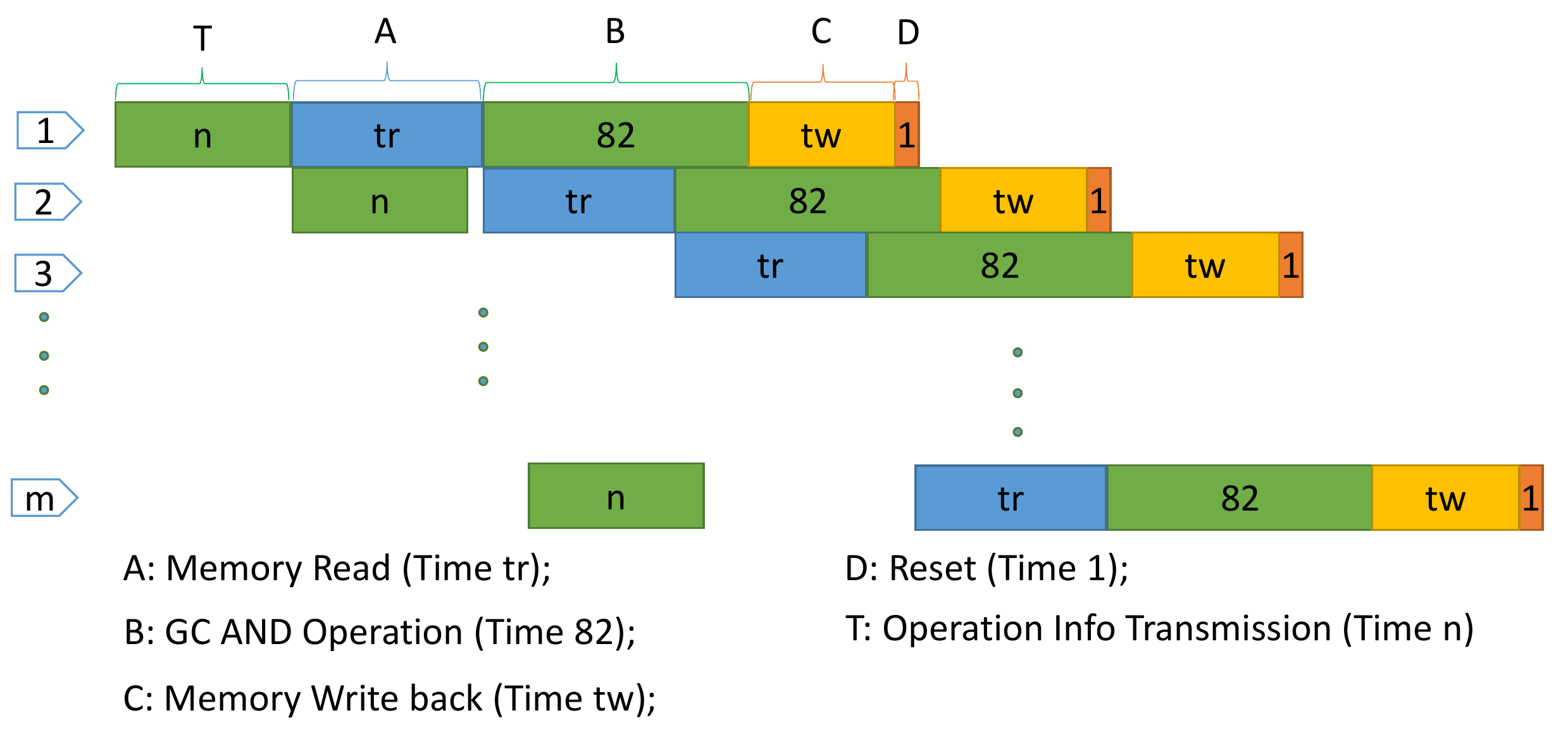}
\end{center}
\caption{Overlapping CPU and FPGA Communication with FPGA Computation}
\label{fig:finegra}
\end{figure}

Our first optimization involves overlapping communication and computation of gate and wire information, as shown in Fig.~\ref{fig:finegra}.  Overlay cells can start working as soon as the information for a new Boolean operation has been transmitted.   For different batches, the same gates implemented as part of the overlay architecture are reused for different garbled gates in the user design.  This optimization is applied in all subsequent designs and in all reported results.  Another optimization we apply is to remove unnecessary handshaking signals between the host and the FPGA. 

The communication channel between the host and FPGA 
supports direct communication to data registers on the FPGA or, using DMA, to on-board DDR memory.
We use DMA to transmit the initial data (values on input wires) to DDR memory. We directly transfer gate information to on-chip registers.  The time for the host to write to one register on the FPGA is 50 ns.  As there are three addresses for a Boolean gate, the data transmission time is 150 ns per gate in a batch. 

\begin{figure}[htb]
\begin{center}
\includegraphics[width=8 cm]{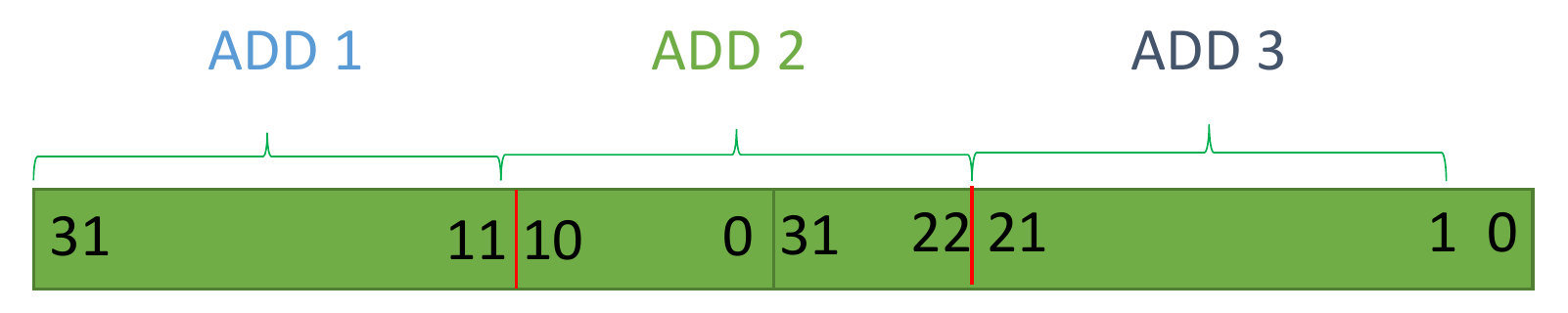}
\end{center}
\caption{Reducing Number of Registers}
\label{fig:3to2}
\end{figure}

An optimization we apply is to pack more than one address into a register to reduce the number of transfers required.  We use two registers to represent the three addresses needed for each gate. The total width of 2 addresses is $2 \; registers*32 \; bits/register = 64 \; bits$ and the actually bit-width for each address location in our design is $\lfloor 64/3 \rfloor = 21$ bits. Besides the flag bit representing the memory type, there will be 20 bits for a real address,  which is enough to represent about one million wires.  Fig.~\ref{fig:3to2} shows this optimization. We are investigating generating memory addresses locally to the FPGA, which will remove this limitation.    

\subsubsection{Hybrid Memory Hierarchy}

The second source of bottleneck in our design is transfers between the FPGA and the on-board DDR memory.  Accesses to DDR memory require many clock cycles with a latency of about 180ns, and, as wires are not accessed in sequential order, we cannot take advantage of burst mode.  Block RAM available on the FPGA has much faster access times of one clock cycle (5ns for a 200MHz CLK), but is not large enough to support the size of problems we are processing.  To address this issue, we make use of a hybrid memory hierarchy where some values are stored in on-chip memory while most values are stored off-chip.  
In essence, we are using the on-chip memory as a cache.  However, unlike a traditional cache, the policy for using the cache is completely under user control.  

Fig. \ref{fig:hybridmem} shows the hardware architecture using both the block RAM (BRAM) inside the FPGA and the DDR memory on board. Our previous work only used on-chip BRAM and thus was limited in the size of problems garbled~\cite{fang2017secure}.
In this research we investigated two different allocation policies, for block RAM.  We refer to these as directly-used and most-frequently-used.  Results for both policies are reported in Sec.~\ref{chap:results}.  Software on the host determines whether a wire is stored in block RAM or in DDR.  A single bit in the address indicates which it is.  Wire IDs are generated on the host, so the code that generates wire IDs also implements the memory policy.  Using a bit to indicate the location reduces the number of addressable memory locations, but also removes the need to implement hardware to track locations of specific locations.  

Addresses for wires are used to store values that represent the output of the garbled gate.  These values are used in generating the garbled values for the next gate; however they are not transmitted to the evaluator.  Only the garble tables need to be transmitted.  Hence values are stored in memory for the duration of the garbling computation across all layers, but are not needed after that. 

Some wires are generated as outputs from one gate and feed directly into another gate.  In other words, their fanout is 1.  In the directly-used policy we store the values generated on wires that are directly used in block RAM to save the time to store and fetch these values.   The criteria for such a value to be stored in block RAM are:  (1) the wire is used only once after it is generated and (2) the Boolean gate which uses this wire ID is in the adjacent layer. 
 The directly used policy saves significant memory bandwidth. The second criteria of only using values in an adjacent layer allows block RAM space to be reused once the garbled value is no longer needed.  For the directly-used policy we use a ping-pong buffering approach, where half the block RAM in any layer is used for reading and the other half for writing, and these roles are swapped with each layer.  A total of 13 Megabits of on-chip BRAM is used.  

For the most-frequently-used policy, the host code analyzes the complete netlist that is generated, and identifies the wires that are most frequently used.  These values are stored in block RAM for the duration of their lifetime.  This policy is similar to a most frequently used cache allocation policy; however in our design, the values are never stored in DDR RAM.  The goal behind this policy is to reduce the number of reads and writes to off-chip memory.  
The host code sorts wire IDs based on their number of accesses and assigns those wires with a large number of accesses to  block RAM.   Once block RAM cannot fit more wires, the rest of the wire IDs are assigned to addresses in DDR Memory.  We do not currently reuse memory locations; this is planned for future work.  
\begin{table}[htb]
\centering
\begin{scriptsize}
\caption{Wire Information for Problems}{
\begin{tabular}{lccccccc}
\hline
Problem & Wire & A Wire  & B Gate & C Gate & Max D & Wire/Layer \\
\hline
6-bit adder  & 42  & 12   & 12 & 0  & 1      & 2.5         \\
\hline
10-bit HD     & 140     & 55         & 50           & 5     & 7    & 6.4         \\
\hline
30-bit HD           & 420     & 163                                 & 147                                           & 11                                              & 22                             & 15.6        \\
\hline
50-bit HD           & 700     & 293                                  & 269                                           & 24                                              & 37                             & 21.9        \\
\hline
8-bit mult         & 495     & 296                                 & 247                                           & 49                                              & 64                             & 8.7         \\
\hline
16-bit mult        & 2015    & 1232                                 & 1007                                          & 225                                             & 256                            & 16.7        \\
\hline
32-bit mult        & 8127    & 5024                                  & 4063                                          & 961                                             & 1024                           & 32.6        \\
\hline
64-bit mult        & 32639   & 20288   & 16319       & 3969     & 4096                           & 64.6        \\
\hline
10 4-bit sorting    & 5717    & 2968                                & 2136                                          & 832                                             & 40                             & 20.6        \\
\hline
$ 5 \times 5$ 4-bit m\_mult   & 16175   & 9700                               & 8350                                          & 1350                                            & 2000                           & 647.0       \\
\hline
$10 \times 10$ 10 10 4-bit m\_mult & 31472   & 18768                               & 16051                                         & 2717                                            & 3809                           & 1165.6      \\
\hline
$5 \times 5$ 8-bit m\_mult   & 64375   & 39400                                & 32850                                         & 6550                                            & 8000                           & 1129.4      \\
\hline
$10 \times 10$ 8-bit m\_mult & 517500  & 317600                               & 263400
                                       & 54200
                                        & 64000                          & 9078.9      \\
\hline
$20 \times 20$ 4-bit m\_mult & 1050800 & 635200                              & 541600                                              &  93600                  & 128000                         & 28400.0    \\
\hline
\end{tabular}}
\label{wireinfo}

A: 1-to-1 wire; \\
B: gate with one 1-to-1 wire from adjacent layer; \\
C: gate with one 1-to-1 wire NOT from adjacent layer; \\
D: Max number of 1-to-1 wires in a layer.
\end{scriptsize}
\end{table}

Table \ref{wireinfo} shows wire information. Each wire corresponds to one memory location. The more wires, the more memory needed to store the values that correspond to the output of each garbled gate. The problems have a wide range of number of wires from several dozen to over one million wires. There are two types of wires, 1-to-1 wires and 1-to-N wires. 1-to-1 wires include two types: wires where the output is immediately used in the next layer, and wires where the output is not immediately used. We use the first type in the directly-used policy.   There are also 1-to-1 wires  not in adjacent layers and 1-to-n wires where one output is used multiple times.  The maximum number of 1-to-1 wires in a layer is the number of memory locations needed in Block RAM for directly-used policy if all 1-to-1 wires are kept on chip.

\begin{table}
\centering
\begin{scriptsize}
\caption{Wire Percent for Problems}{
\begin{tabular}{lccc}
\hline
Problem             & Percent A & Percent B & Percent C \\
\hline
6-bit adder         & 28.57\%            & 100.0\%                              & 28.6\%                         \\
\hline
10-bit HD           & 39.29\%            & 90.9\%                               & 35.7\%                         \\
\hline
30-bit HD           & 35.00\%            & 90.2\%                              & 38.8\%                         \\
\hline
50-bit HD           & 41.86\%            & 91.8\%                               & 38.4\%                         \\
\hline
8-bit mult         & 59.80\%            & 83.4\%                               & 49.9\%                         \\
\hline
16-bit mult        & 61.14\%            & 81.7\%                               & 50.0\%                         \\
\hline
32-bit mult        & 61.82\%            & 80.9\%                               & 50.0\%                         \\
\hline
64-bit mult        & 62.16\%            & 80.4\%                               & 50.0\%                         \\
\hline
10 4-bit sorting    & 51.92\%            & 72.0\%                               & 37.4\%                         \\
\hline
$5 \times 5$ 4-bit m\_mult   & 59.97\%            & 86.1\%                               & 51.6\%                         \\
\hline
$10 \times 10$ 4-bit m\_mult & 59.63\%            & 85.5\%                               & 51.0\%                         \\
\hline
$5 \times 5$ 8-bit m\_mult   & 61.20\%            & 83.4\%                               & 51.0\%                         \\
\hline
$10 \times 10$ 8-bit m\_mult & 61.37\%            & 82.9\%                               & 50.9\%                         \\
\hline
$20 \times 20$ 4-bit m\_mult & 60.45\%            & 85.3\%                               & 51.5\%                          \\
\hline
\\
\end{tabular}}
\label{wireperc}

Percent A: Percent of 1-to-1 wire in all wires; \\Percent B: 1-to-1 wires to be used in the next layer of all 1-to-1 wires; \\Percent C: 1-to-1 wires to be used in the next layer of all wires
\end{scriptsize}
\end{table}

Table \ref{wireperc} shows the percentage of each type of wire. Percent A is the number of 1-to-1 wires among all the wires. Most of the 1-to-1 wires are used in the next layer, represented in Percent B. Percent C shows the percent of the 1-to-1 wire to be used in the next layer among all the wires.  These data show that the directly-used policy is a good fit for many garbled circuit problems, and may require less on-chip memory  compared to the most-frequently-used policy. 

For the hardware architecture, the workload dispatcher and data controller is designed to accommodate this hybrid memory hardware architecture. The controller monitors the flag of the address provided by the host, and determines whether the value should be stored in  Block RAM or DDR. We embed the flag as the last bit of the address, and if zero, the location is DDR; otherwise it is block RAM.  Timing results for both of the implemented policies are presented in the next section.

\section{Experiments and Results}
\label{chap:results}

We compare our results to FlexSC~\cite{flexsc} both for correctness and for performance.  
For software timing, we run FlexSC on an Intel Core i7 processor running at 3.6GHz, using any optimizations implemented in FlexSC. 
Our target hardware platform consists of a host PC and FPGA card, specifically the ProceV board from Gidel. The ProceV board hosts a Stratix V FPGA with two DDR3 external memories each of which can support 8GB, or a total of 16GB. It provides communication between host and FPGA via a PCIe Gen 3 bus with 8 lanes, each of which supports 8 Gigatransfers per second.  Our results show the end-to-end effect of replacing FlexSC garbling in software with an FPGA solution.  

In our results, we examine the effect of the basic overlay architecture that we implemented, as well as each of the optimizations discussed in Sec.~\ref{arch}. Note that all of the results presented use overlapped communications and computation as described in Sec.~\ref{pcieorg}.  In the reported results, the FPGA clock is 200 MHz. The interface clock responsible for data transmission between host and FPGA is running at 300MHz.  Timing results compare the FPGA design to an Intel processor running the same algorithm at 3.6 GHz.

\subsection{Problem Analysis}
\label{analysis}

The problems we analyze are addition, Hamming Distance (HD), multiplication, sorting,  and matrix multiplication. We also analyze scalability of our design by testing several different sizes of these problems.  Results for some of these problems were reported in our previous publication~\cite{fang2017secure}.  The biggest difference between the work presented here and our previous work is that the previous work did not make use of off-chip memory, which significantly limited the size of problems we can garble. The maximum number of gates in a previously presented example was 32,000 gates.  Here we analyze much larger problems, with up to a million gates.  Problem information is summarized in Table~\ref{gateinfo-motiv}.  This table shows total numbers of layers, wires, and gates in the garbled circuit problems we analyze.  The examples presented are a 6-bit adder, several different bit sizes for HD, several different bit sizes for multiplication, and an example of sorting four bit numbers.  All of these were reported previously~\cite{fang2017secure}.  The new examples include different sizes of matrix multiplication that show how the problem size scales as well as larger problems that could not fit with the previous approach.  The largest problem reaches one million garbled circuit operations and several thousand independent gates within each layer. Note that the largest number of layers is not from the largest problem.  Also note that, to best take advantage of the free XOR optimization, FlexSC generates examples that  predominantly make use of gXORs; gANDs never exceed more than 26\% of total gates.  In addition to these examples, we analyze different sizes of page rank (PR) and present results below.  

\subsection{Heterogeneous Computing System Results}
\label{hetresult}

We have implemented all of the design variants described in this paper on a Gidel ProceV board.  In this section we present step by step performance improvements using the different optimizations described is Sec.~\ref{arch}.  We compare subsequent designs to one another, and also present the speedup compared to software using FlexSC. 

\begin{table}[htb]
\centering
\begin{scriptsize}
\caption{Increase Number of AND Overlay Cells}{
\begin{tabular}{lcccc}
\hline
Problem & 5 AND Overlay (us) & Speed-up & 10 AND Overlay (us) & Total Speed-up \\ \hline
6-bit adder & 78 & 26.41 & 76 & 27.11 \\ \hline
10-bit HD & 260 & 9.73 & 257 & 9.84 \\ \hline
30-bit HD & 765 & 5.33 & 741 & 5.51 \\ \hline
50-bit HD & 1282 & 5.04 & 1210 & 5.34 \\ \hline
8-bit mult & 1098 & 8.40 & 1058 & 8.71 \\ \hline
16-bit mult & 4280 & 3.40 & 4218 & 3.45 \\ \hline
32-bit mult & 17406 & 1.94 & 17056 & 1.98 \\ \hline
64-bit mult & 71068 & 2.15 & 69858 & 2.19 \\ \hline
10 4-bit sorting & 12605 & 1.68 & 12375 & 1.71 \\ \hline
\end{tabular}}
\label{numcells}
\end{scriptsize}
\end{table}

We experiment with different numbers of overlay cells implemented in hardware, as shown in Table~\ref{numcells}.   Results show speedup compared to FlexSC for 5 and 10 garbled AND gates; in both cases a single XOR overlay cell is used.  We do not observe much improvement when we increase the number of AND gates, which indicates that this is not the bottleneck in our base design.  The bottleneck here is PCIe communications between host and FPGA board.  

\begin{table}
\centering
\begin{scriptsize}
\caption{Results for Removing Host XOR Operation Check}{
\begin{tabular}{lccc}
\hline
Problem          & 10 AND w/o xor check (us) & Additional Speed-up & Total Speed-up \\ \hline
6-bit adder      & 60                   & 1.30                  & 34.33          \\ \hline
10-bit HD        & 99                   & 2.63                  & 25.56          \\ \hline
30-bit HD        & 216                  & 3.43                  & 18.89          \\ \hline
50-bit HD        & 365                  & 3.32                  & 17.70          \\ \hline
8-bit mult      & 428                  & 2.47                  & 21.54          \\ \hline
16-bit mult     & 1420                 & 2.97                  & 10.24          \\ \hline
32-bit mult     & 4924                 & 3.46                  & 6.86           \\ \hline
64-bit mult     & 18673                & 3.74                  & 8.20           \\ \hline
10 4-bit sorting & 2770                 & 4.47                  & 7.62           \\ \hline
\end{tabular}}
\label{remxor}
\end{scriptsize}
\end{table}

In the base design, the transmission time for information for each XOR operation is larger than the garbling XOR time itself.  Using this information, we can remove any synchronization used by the host when sending XOR gates.   Note that it is still worthwhile to garble the XOR gates on the FPGA, because the output keys generated are used in subsequent gates in the design.  
Knowing that transmission time of each XOR operation is larger than the XOR operation time, we can remove the synchronization steps and let the host keep sending XOR gates. Table~\ref{remxor} shows the results of applying this optimization to the ten garbled AND gate design. Additional speed-up is the speed-up compared with the version with xor check and total speed-up compares this new design with FlexSC. In this design, the host sends all the XOR operations within one layer without synchronization before sending batches of AND operations. This optimization contributes significant speedup, and the effect of this optimization grows as the size of the user problem increases.  

\begin{table}[htb]
\centering
\begin{scriptsize}
\caption{Directly-Used Policy using block RAM and DDR Hybrid Memory}{
\begin{tabular}{lcc}
\hline
Problem & 10AND + Hybrid Memory (us)& Speed-up \\ \hline
6-bit adder & 54 & 57.2 \\ \hline
10-bit HD & 88 & 28.8 \\ \hline
30-bit HD & 193 & 21.1 \\ \hline
50-bit HD & 302 & 21.4 \\ \hline
8-bit mult & 380 & 24.3 \\ \hline
16-bit mult & 1284 & 11.3 \\ \hline
32-bit mult & 4208 & 8 \\ \hline
64-bit mult & 15945 & 9.6 \\ \hline
10 4-bit sorting & 2292 & 9.2 \\ \hline
\end{tabular}}
\\
\label{dup2}
Hybrid Memory consisting of block RAM on FPGA and DDR on Board
\end{scriptsize}
\end{table}

Next, we show the speedup from using both on-chip Block RAM and off-chip DDR memory.  
Table \ref{dup2} shows total time in $\mu$s and the speedup compared to the software version in FlexSC.  In this table, we apply the directly-used policy, described in Sec.~\ref{optim}.  The results show that the smallest speedup compared to software is 8 times.  Thus using a hybrid memory architecture results in significant savings.

\begin{table}[htb]
\centering
\begin{scriptsize}
\caption{Most-Frequently-Used Policy}{
\begin{tabular}{lcc}
\hline
Problem & 10AND + Hybrid Memory 2 (us) & Policy Comparison \\ \hline
32-bit mult & 4384 & 1.04 \\ \hline
64-bit mult & 15648 & 0.98 \\ \hline
10 4-bit sorting & 2425 & 1.06 \\ \hline
\end{tabular}}
\label{mfup}
\end{scriptsize}
\end{table}

We also implemented the most frequently used policy using the hybrid memory system. We try three of the larger problems to compare the two policies, as shown in Table~\ref{mfup}. The policy comparison column compares the most frequently used policy with the directly used policy.  If it is larger than 1, then the most frequently used policy is faster. Note that the results show that there is not much difference in performance between the two policies for large problems.  This is likely due to the fact that the directly used policy reuses block RAM memory locations,  while in our current implementation, the  most frequently used policy does not.  
In addition, the pre-processing cost for the most frequently used policy is more expensive as the fanout of every wire needs to be computed.  In addition, to reuse memory locations, lifetimes of these wires will need to be computed.  While we plan to investigate this in the future, for now, we conclude that the directly used policy is the most advantageous.   

\begin{table}[htb]
\centering
\begin{scriptsize}
\caption{Influence of Number of Gates}{
\begin{tabular}{cccc}
\hline
Gates Number & Time & Speedup Compared with SW & Speedup Improvement \\ \hline
5XOR 5AND & 18677 & 8.20 & - \\ \hline
10XOR 10AND & 14888 & 10.29 & 1.25 \\ \hline
15XOR 15AND & 12252 & 12.50 & 1.22 \\ \hline
\end{tabular}}
\label{numgates}
\\64-bit multiplication problem. 300 MHz main clock and 200 MHz local clock.
\end{scriptsize}
\end{table}

For the FPGA operation we use 200MHz as the local clock.  The PCIe protocol allows us to set a different ``main'' clock speed for transmitting data; for this we use 300Mz.  Because the main clock is faster, the time to transmit the operands for a garbled XOR is no longer larger than the XOR operation time. Thus, we can not apply XOR without synchronization between the host and FPGA. However, we can also use multiple XORs to improve the total performance. Table~\ref{numgates} shows the results of using 5 AND and 5 XOR overlay cells; 10 AND and 10 XOR; 15 AND and 15 XOR. Speedup improvement shows that the increase from changing from 5 to 10 is 1.25 times and changing from 10 to 15 is 1.22 times.  We will continue to investigate adding more gates to see when this improvement saturates.  

\begin{table}[htb]
\centering
\begin{scriptsize}
\caption{Using 2 Address Registers for 3 Addresses}{
\begin{tabular}{ccccc}
\hline
Problem & 1 Reg as 1 address & 2 Regs as 3 addresses & Improvement & Total Speedup\\ \hline
2 PR & 41044 & 37358 & 1.1 & 12.47 \\ \hline
3 PR & 66409 & 58587 & 1.13 & 10.27 \\ \hline
4 PR & 90087 & 7.83 & 1.06 & 7.83 \\ \hline
\end{tabular}}
\label{2reg3add}
\\10AND and 10XOR overlay cells; 300 MHz main clock and 200 MHz local clock.  PR is Page Rank.  
\end{scriptsize}
\end{table}

The speedup results from packing 3 addresses into 2 registers are shown in Table~\ref{2reg3add}. We use the page-ranking examples and the results show 1.06 to 1.13 speedup improvement compared with the method of using 1 register for 1 address.  Note that this optimization limits the size of valid address bits that can be used to 20, which in turn limits the size of problems that can be garbled. 

\begin{table}[htb]
\centering
\begin{scriptsize}
\caption{Speedup Results}{
\begin{tabular}{lccc}
\hline
Problem & sw (ms) & Time (us) & Speedup \\ \hline
6-bit adder & 2.06 & 45 & 45.78 \\ \hline
10-bit HD & 2.53 & 80 & 31.63 \\ \hline
30-bit HD & 4.08 & 171 & 23.86 \\ \hline
50-bit HD & 6.46 & 259 & 24.94 \\ \hline
8-bit mult & 9.22 & 293 & 31.47 \\ \hline
16-bit mult & 14.54 & 949 & 15.32 \\ \hline
32-bit mult & 33.76 & 3308 & 10.21 \\ \hline
64-bit mult & 153.13 & 12252 & 12.50 \\ \hline
10 4-bit sort & 21.12 & 2339 & 9.03 \\ \hline
$5 \times 5$ 4-bit m\_mult & 60.66 & 5830 & 10.40 \\ \hline
$10 \times 10$ 4-bit m\_mult & 220.81 & 11286 & 19.56 \\ \hline
$5 \times 5$ 8-bit m\_mult & 203.86 & 24128 & 8.45 \\ \hline
$10 \times 10$ 8-bit m\_mult & 1060.63 & 170895 & 6.21 \\ \hline
$20 \times 20$ 4-bit m\_mult & 2170.88 & 340698 & 6.37 \\ \hline
\end{tabular}}
\label{final}
\end{scriptsize}
\end{table}

We combine all of the optimizations that led to speedup and present the results in Table~\ref{final}.  These results have applied the following optimizations: (1) 15 AND overlay cells and 15 XOR overlay cells; (2) Hybrid memory system with the directly-used policy; (3) 300 MHz main clock frequency for PCIe interface and 200 MHz local clock frequency; (4) Pipelined operation between the host and FPGA.  The results are shown for working designs on the Gidel ProceV board and compare end-to-end system running time to the same problems run in software using FlexSC.  We observe one or two orders of magnitude speedup across a range of problems.  Note that software is running at 3.6GHz, while the FPGA implementations are running at 200 MHz.  FlexSC runs with one thread; however parallelizing the particular implementation of GC with the optimizations  used and the ``honest but curious'' model is not trivial. Note that the number of AND gates garbled per second continues to increase as the size of the problem grows.  While we see significant speedup across all problems, the amount of speedup diminishes as the problem size grows.  This is due to the fact that on chip BRAM cannot keep as large a percent of memory locations off of off-chip memory as the size of each layer grows and highlight the importance of our hybrid memory optimization.  We intend to continue to optimize our design to be able to garble larger and larger problems in less elapsed time.  

\subsection{Bandwidth Bottleneck}

 Under the two optimizations we employ (``free'' XOR
gates \cite{Kolesnikov2008}  and garbled-row reduction \cite{PSSW09}), the garbler needs to send $3 \times 80$-bit ciphertexts ($240$ bits) to the evaluator \emph{per AND gate}, and 0 ciphertexts (0 bits) \emph{per XOR gate} in the circuit. The latter is due precisely to the use of the ``free'' XOR optimization: garbled XOR gates require neither encryption during garbling  nor any transmission  using this technique.

Under these optimizations, garbling is computation-bound in our setting. Taking two cases in Table~\ref{final} ($10 \times 10$ 4-bit m\_mult, $20 \times 20$ 4-bit m\_mult) as examples, our processing time indicates we can process $0.67$M and $0.75$M AND gates per second, respectively. These cases correspond to the case with the largest speedup and the largest example run.  At the cost of 240 bits per gate, this garbling correspond to a required communication bandwith between garbler and evaluator of 160.8MBps and 180Mbps, respectively. This is well within the range of the bandwidth available at, e.g., Amazon Web Services (AWS) EC2 instances (5GBps); this implies that a garbler and evaluator deployed by distinct entities on AWS would be computation, not communication-bound. We note that this observation, as well as our estimates, agree with experimental observations of garbler-evaluator execution pairs on AWS \cite{nayak2015graphsc}.

\section{Conclusions and Future work}
\label{chap:conclusion}

This article demonstrates a heterogeneous reconfigurable computing system using FPGA overlay architecture for general garbled circuit operations. 
This system lets the user implement and accelerate their application without any knowledge of either hardware development or secure function evaluation protocol by providing a complete workflow to transfer {\em any} garbled circuit problem onto it. We demonstrate the benefit of using this system by showing significant speedup compared with existing software platforms. This research makes possible the wider adoption of using garbled circuit schemes in the future.

For the hardware architecture on FPGA, our design uses a coarse-grained overlay architecture and enables the evaluation of different SFE tasks without the need for reprogramming. The host side workflow includes garbled circuit generator, problem parser, and host code generation tools which can be configurable for different hardware architectures. These tools explore the parallelism for any GC problem and generate the host program based on the structure of the problem. We also provide analytical tools to show the different characteristics of a problem.
We explore the bottlenecks while working on this heterogeneous reconfigurable computing system and tackle them using different methods. This exploration also provides other researchers directions for improving their own heterogeneous system designs.

There are several directions for future research.  First is the further improvement of the heterogeneous system.  This research may benefit from  a closely connected FPGA such as the Intel HARP to alleviate the bottleneck of the PCIe interface.  Another direction is to expand the overlay cell library to abstract more complicated computational patterns using Boolean AND and XOR operations. The current work uses garbled circuit AND and XOR overlay cells as two components of the hardware architecture library, and this fine-grained pattern suffers from DDR access delay in every operation. Based on Tables \ref{wireinfo} and \ref{wireperc}, we know that there are many 1-to-1 wires to be used in the next layer. One solution is to build other overlay cells which consist of cross-layer Boolean operations.
Second is to separate a large problem into several small problems which can be computed independently through several host nodes each with its own  FPGA board. This enables the expansion of the size of the problems into even larger data mining problems, such as page ranking with more nodes using GraphSC and eventually provide a large, scalable, efficient platform for privacy-preserving computation.  We have already begun to test these ideas using Amazon Web Services F1 instances.

\section*{Data Availability} The examples used to support the findings of this study are available from the corresponding author upon request.

\section*{Acknowledgements}

This material is based upon work supported by the National Science Foundation under Grant No. 1717213.  The research was also supported by a Google Faculty Research Award.  We would like to thank Mehmet Gungor and Kai Hwang for valuable contributions to this research.  

The authors declare that there is no conflict of interest regarding the publication of this paper.

\bibliographystyle{unsrt}
\bibliography{thesis}

\end{document}